\documentclass[%
 reprint,
 amsmath,amssymb,
 aps,prl,superscriptaddress
]{revtex4-1}

\usepackage{graphicx}% Include figure files
\usepackage{dcolumn}% Align table columns on decimal point
\usepackage{bm}% bold math
\usepackage[mathlines]{lineno}% Enable numbering of text and display math
\begin{document}

\preprint{APS/123-QED}

\title{Measurement of the $^{12}$C($\boldsymbol{n,p}$)$^{12}$B cross section at n\_TOF (CERN) by \textit{in-beam} activation analysis}

\author{P.~\v{Z}ugec} \affiliation{Department of Physics, Faculty of Science, University of Zagreb, Croatia}
\author{N.~Colonna} \email[]{nicola.colonna@ba.infn.it}
\affiliation{Istituto Nazionale di Fisica Nucleare, Sez. di Bari, Italy}
\author{D.~Bosnar} \affiliation{Department of Physics, Faculty of Science, University of Zagreb, Croatia}
\author{A.~Mengoni}\affiliation{ENEA, Bologna, Italy}
\author{S.~Altstadt} \affiliation{Johann-Wolfgang-Goethe Universit\"{a}t, Frankfurt, Germany}
\author{J.~Andrzejewski} \affiliation{Uniwersytet \L\'{o}dzki, Lodz, Poland}
\author{L.~Audouin} \affiliation{Centre National de la Recherche Scientifique/IN2P3 - IPN, Orsay, France}
\author{M.~Barbagallo} \affiliation{Istituto Nazionale di Fisica Nucleare, Sez. di Bari, Italy}
\author{V.~B\'{e}cares}  \affiliation{Centro de Investigaciones Energeticas Medioambientales y Tecnol\'{o}gicas (CIEMAT), Madrid, Spain}
\author{F.~Be\v{c}v\'{a}\v{r}} \affiliation{Charles University, Prague, Czech Republic}
% ...

\collaboration{The n\_TOF Collaboration (www.cern.ch/ntof)}  \noaffiliation

\date{\today}

\begin{abstract}
The integral cross section of the $^{12}$C($n,p$)$^{12}$B reaction has been determined for the first time in the neutron energy range from threshold to several GeV at the n\_TOF facility at CERN. The measurement relies on the activation technique, with the $\beta$-decay of $^{12}$B measured over a period of four half-lives within the same neutron bunch in which the reaction occurs. The results indicate that model predictions, used in a variety of applications, are mostly inadequate. The value of the integral cross section reported here can be used as a benchmark for verifying or tuning model calculations.

\begin{description}
\item[PACS numbers]
\pacs{}23.40.-s, 24.10.Lx, 28.20.Fc

\keywords{Prove tecniche}

\end{description}

\end{abstract}

\maketitle

%\tableofcontents

Neutron cross-section data are important for several fields of fundamental and applied Nuclear Physics. In particular, cross sections for neutron-induced reactions with carbon, oxygen, nitrogen and other light elements abundantly present in the human body are needed in order to accurately estimate the dose to tissues in treatments with neutrons as well as in radiotherapy with protons and light ions. Particularly significant in this respect are the reactions leading to the emission of charged particles. Among them, the $^{12}$C($n,p$)$^{12}$B reaction, occurring at neutron energies above the reaction threshold of 13.6~MeV, may affect the dose distribution in hadrontherapy or conventional radiotherapy in the presence of a high-energy neutron field. Together with protons, electrons with an average energy of 6.35~MeV are emitted as a consequence of the decay of $^{12}$B, characterized by a very short half-life of 20.2~ms \cite{ajzen}. The ($n,p$) cross section for carbon is also a basic input in calculations of radiological protection, as well as for the design of shields and collimators at accelerator-based neutron facilities, in particular spallation neutron sources and fusion material irradiation facilities, such as MTS and IFMIF, whose neutron spectrum presents an important tail extending above the threshold of this reaction \cite{fusion}. Other applications of high-energy neutron beams, as for example Accelerator Driven Systems, may also benefit from new data on this reaction. Finally, given the increasing importance of diamond detectors, new cross section data on the $^{12}$C($n,p$)$^{12}$B reaction would be desirable, to improve simulations of the detector response to fast neutrons \cite{rebai,pillon}.

\begin{figure}[b]
\includegraphics[angle=90,width=1.\linewidth,keepaspectratio]{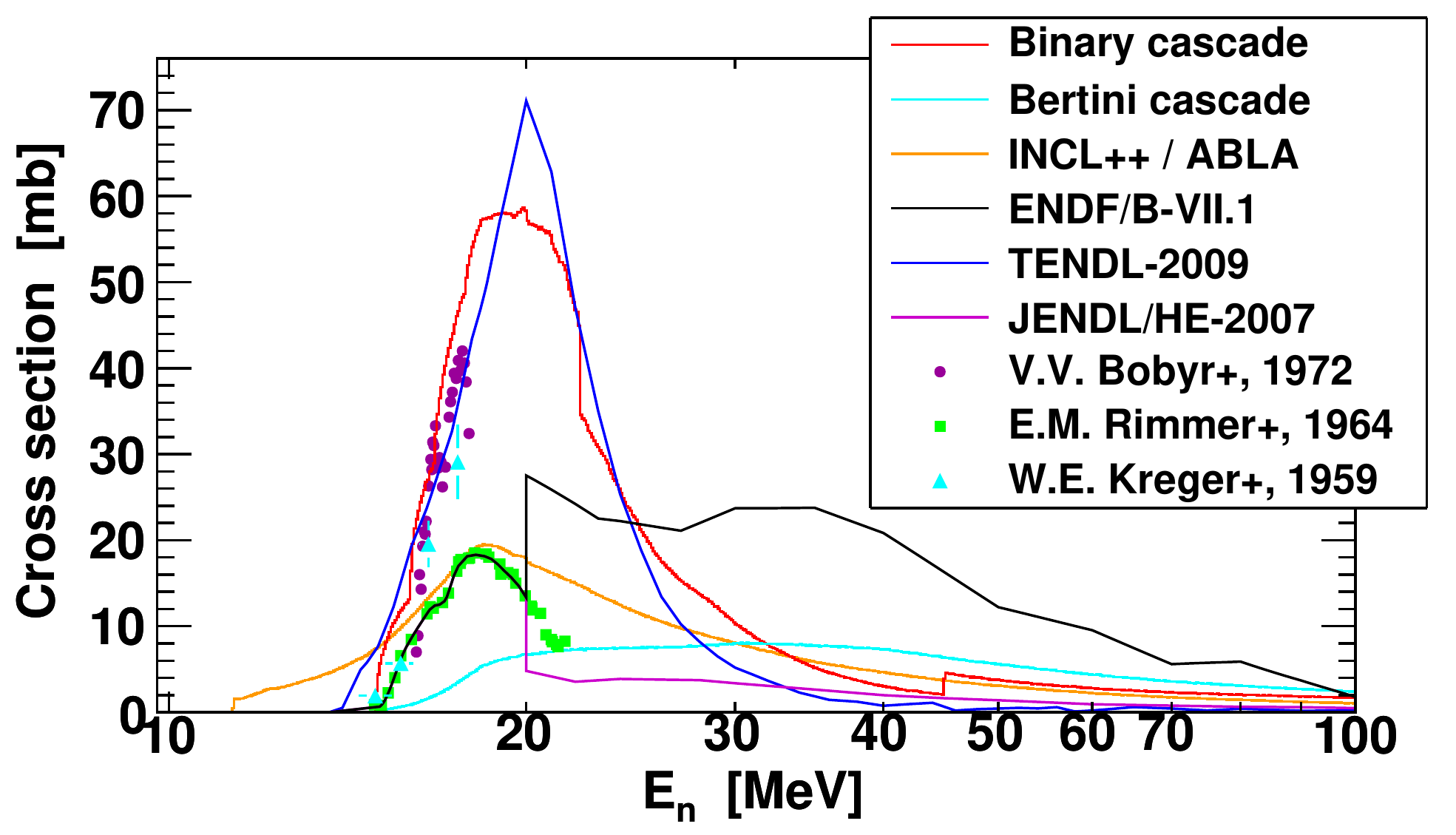}
\caption{\label{fig:one} (Color online) Current status of the $^{12}$C($n,p$)$^{12}$B cross section. Symbols represent experimental data, while lines show evaluated cross sections and model calculations.}
\end{figure}

At present, cross section data on this reaction are scarce and largely discrepant. Fig.~\ref{fig:one} shows the current status of the cross sections. Only three datasets are reported in literature, extending only a few MeV above threshold \cite{kreger,rimmer,bobyr}. Two of them \cite{kreger,rimmer} were obtained by means of the activation technique, with short pulses of monoenergetic neutrons inducing the reaction, followed by long beam-off intervals for counting the $^{12}$B $\beta$-decay. The lack of data on this reaction reflects on the evaluated cross section and on model calculations, often used in Monte Carlo codes for neutron transport. Up to 20 MeV, all major evaluated data libraries  contain the same cross section,  based purely on the dataset from Rimmer \textit{et al.} \cite{rimmer}. 
The only exception is TENDL-2009 which, based on TALYS calculations \cite{talys}, predicts a cross section a factor of three higher relative to all other evaluations. Another major problem of the evaluated cross sections is their limited energy range. Above 20~MeV one can only rely on theoretical estimates, such as from the optical model calculations of Ref.~\cite{dimbylow}. Calculations performed with the Feshbach-Kerman-Koonin (FKK)-GNASH code described in Ref.~\cite{chadwick} have been adopted in ENDF/B-VII.1 to extend the cross sections from 20 to 150~MeV \cite{endf}. A completely different cross section, based on calculations by Watanabe \textit{et al.}, \cite{wata} is contained in the special high energy file of the japanese evaluted nuclear data library, JENDL/HE-2007 \cite{jendl, jendl_he}.
Together with evaluated libraries, model calculations are commonly used in modern codes of neutron transport. In Fig.~\ref{fig:one} the predictions of three different models available in GEANT4 \cite{geant4} are also shown: the Binary cascade, the Bertini cascade, and the INCL++/ABLA model (see Ref. \cite{binary} for details). While in principle these predictions can be checked against experimental data below 20~MeV, nothing can be said of the validity of the calculations above this energy, due to the lack of experimental data. A new measurement covering a wide energy range, from threshold to several GeV, would therefore be useful as a benchmark for validating the predictions of model calculations.

Time-of-flight facilities based on spallation neutron sources could in principle be used for measuring the differential cross section in a wide energy range. In practice, however, the measurement is complicated by the presence of other competing reaction channels, in particular elastic and inelastic scattering, ($n,d$) and ($n,\alpha$) reactions \cite{rebai}. A somewhat simpler, yet useful approach, would be to measure the integral cross section by means of the activation technique with a pulsed neutron beam of low repetition rate and with an energy spectrum extending much above the reaction threshold. Both requirements are met by the n\_TOF facility at CERN \cite{carlosEPJ}. The white spectrum extending to $\sim$10 GeV, and a low repetition rate ($\leq$0.8 Hz) offered the unique opportunity to measure for the first time the integral cross section of the $^{12}$C($n,p$)$^{12}$B reaction in a wide range above the reaction threshold. Furthermore, contrary to previous activation measurements of this reaction, at n\_TOF the $\beta$-decay of $^{12}$B is detected \textit{within} the same neutron pulse in which activation takes place, with four half-lives covered by the $\sim$90~ms wide acquisition window used for measuring neutron-induced reactions down to thermal neutron energy, in the measuring station located at 187~m from the spallation source.

The measurement was performed with the experimental setup used in measurements of neutron capture cross sections. A detailed description of the apparatus can be found in  \cite{petarPRC}. Briefly, it is based on two deuterated benzene liquid scintillator detectors (C$_6$$^2$H$_6$, denoted as C$_{6}$D$_{6}$), placed on either side of the neutron beam at a few cm distance from the sample, in the backward direction. The two detectors have different active volumes, with the scintillator contained inside a 0.4~mm thick carbon-fiber cell in one case, and a 1.78~mm thick aluminum cell in the other one (we refer to the two detectors as "FZK" and "Bicron" respectively, since the first one was specifically optimized for n\_TOF at Forschungszentrum Karlsruhe, Germany, while the second one was purchased from Bicron Corporation). The relatively energetic electrons from the $^{12}$B $\beta$-decay (hereafter referred to as $^{12}$B-e$^{-}$) can therefore reach the scintillator volume and deposit therein a large fraction of their initial energy. The acquisition window is started by the proton beam impinging on the spallation target. The prompt signal ($\gamma$-flash) from the spallation target is used as reference for the time calibration. The energy deposited in the detectors was calibrated up to 4.4 MeV with $^{137}$Cs, $^{88}$Y and Am/Be $\gamma$-ray sources. 

A high-purity (99,95\%) $^\mathrm{nat}$C sample of 7.13~g mass and 2~cm diameter was used in the measurement. A chemical analysis performed on the sample excluded contamination by high cross section isotopes. 
In order to extract the $^{12}$C($n,p$)$^{12}$B reaction cross section it is necessary to determine with good accuracy the flux of the neutron beam impinging on the sample in the energy region of interest and the efficiency of the setup to the $^{12}$B-e$^{-}$. Furthermore, all possible sources of background should be identified and subtracted. The energy dependence of the n\_TOF neutron flux has been measured with a few percent uncertainty between 10~MeV and 1~GeV with Parallel Plate Avalanche Counters, by means of the $^{235}$U($n,f$) reaction \cite{massimoEPJ}, and constantly monitored during the measurement. Monte Carlo simulations of the spallation process, normalized at lower energy, are used to extend the neutron flux up to 10 GeV. Since the sample is smaller than the beam, the intercepted fraction has to be considered in the analysis. This was determined by means of the saturated resonance technique \cite{borella}, for the 4.9~eV resonance in the Au($n,\gamma$) cross section, and propagated at higher energies on the basis of the simulated beam profile \cite{carlosEPJ}.

The detection efficiency and the neutron background were determined by means of detailed GEANT4 simulations of the experimental setup. The simulations are described in Ref. \cite{petarNIM}. A realistic software replica of the whole setup, including the walls of the experimental area, was implemented in the simulations, together with the energy resolution of the detectors, determined with $\gamma$-ray sources. The efficiency was estimated as a ratio between the number of electrons produced in the sample and those depositing an amount of energy above a given threshold in the detectors. Since the deposited energy spectrum of $^{12}$B-e$^{-}$ is approximately flat up to 4~MeV, a wide range of thresholds, from 200~keV up to 3.5~MeV, was considered in this work. Second order corrections of the geometrical efficiency are accounted for in the normalization factor extracted separately for the two detectors, from the 4.9~eV saturated resonance of $^{197}$Au.

\begin{figure}[b]
\includegraphics[angle=90,width=1.\linewidth,keepaspectratio]{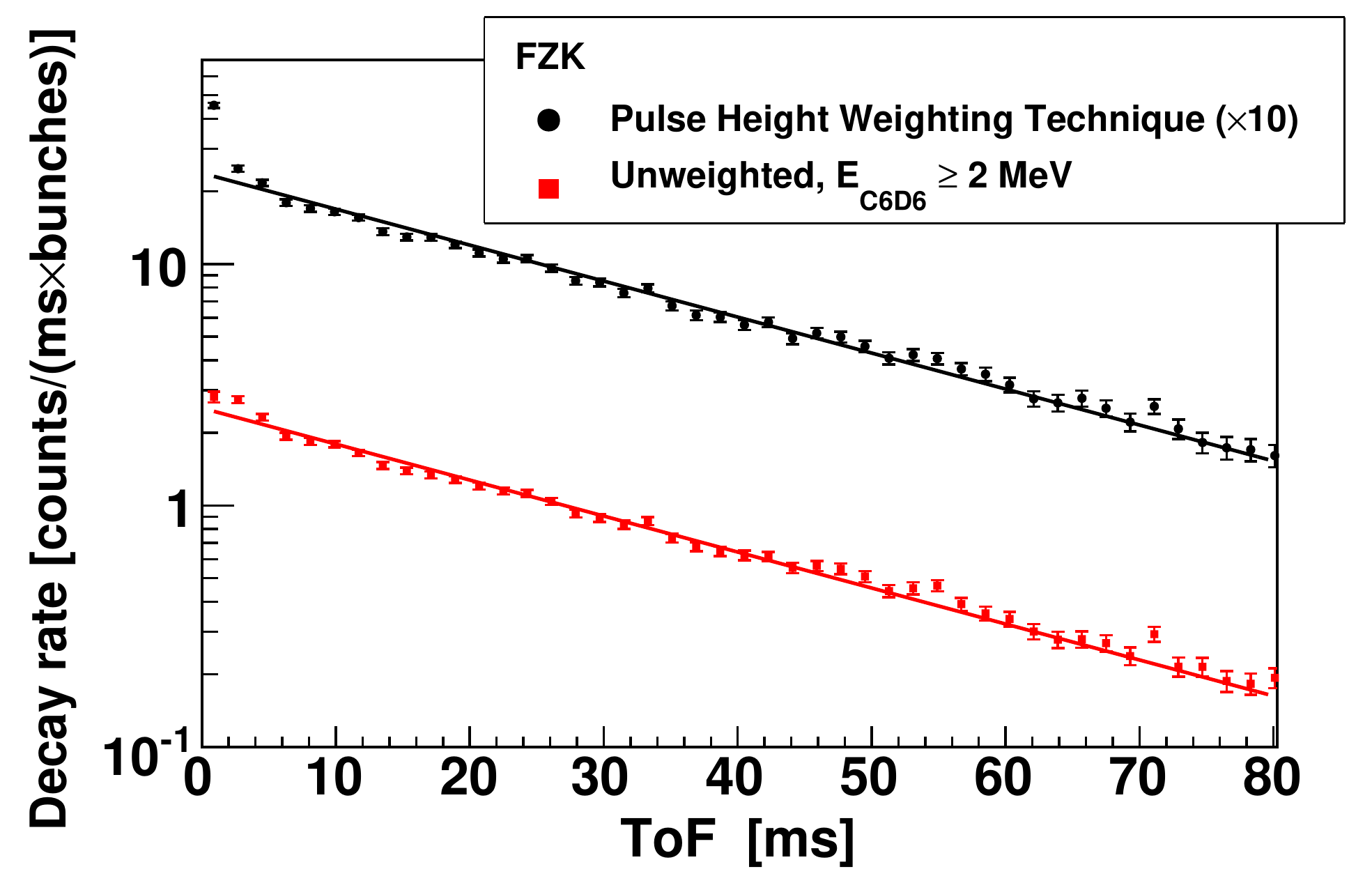}
\caption{\label{fig:two} (Color online) Fit of the time distribution of events in the C$_{6}$D$_{6}$, per neutron bunch, for the two different analysis techniques described in the text. In all cases, data are background-subtracted and corrected for the simulated efficiency. The lines are exponential fits of the data with t$_{1/2}$=20.2~ms.}
\end{figure}

The issue of the neutron-generated background is more complex and needs a careful consideration. Two background components affect the present measurement. The sample-independent background, mostly related to the neutron beam crossing the experimental area, was determined in runs without the sample, and subtracted from the data. The second component, a sample-related one, is produced by neutrons elastically scattered by the sample and subsequently captured in various materials inside the experimental area, including the concrete walls of the hall, with the resulting $\gamma$-rays eventually detected in the C$_{6}$D$_{6}$. Contrary to the sample-independent one, this component cannot easily be measured, and must rely on simulations. A detailed description of this background component can be found in Ref. \cite{petarNIM}. In this work, two different methods have been applied in the analysis of the background, with the consistency of the results checked by comparison.

The first, standard approach, consists in rejecting a large portion of the background by means of a suitable threshold on the deposited energy. Since spurious events are mostly concentrated in the region of low amplitudes, while the deposited energy spectrum of $^{12}$B-e$^{-}$ is nearly flat up to 4~MeV, a high value of the threshold efficienctly suppresses the background, relative to ($n,p$) events. The residual fraction of the background, which includes a small contribution from other radioisotope-producing reactions, is estimated from simulations and subtracted from the data.

\begin{figure}[b!]
\includegraphics[angle=90,width=1.\linewidth,keepaspectratio]{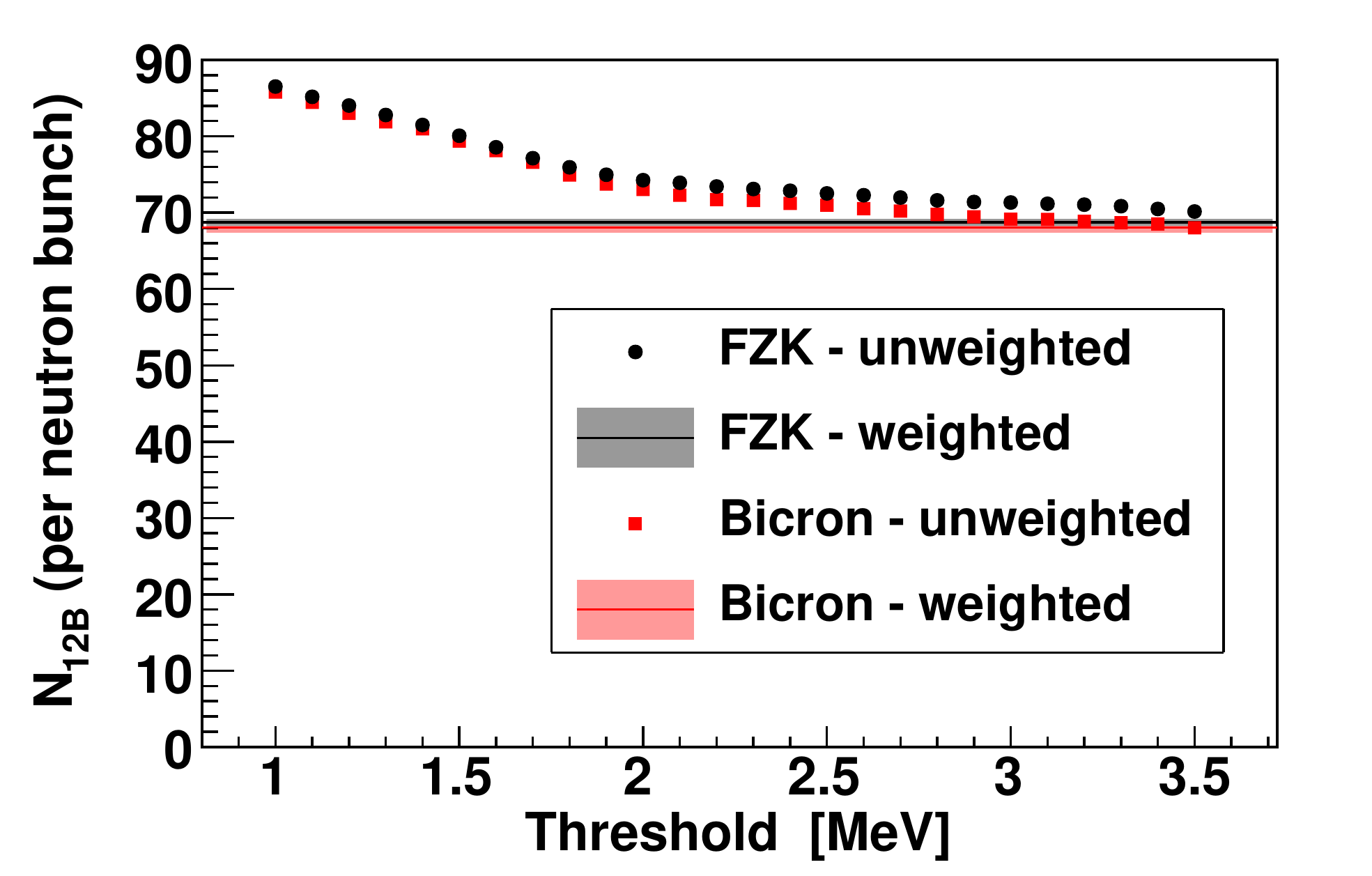}
\caption{\label{fig:three} (Color online) Number of $^{12}$B nuclei produced in the $^\mathrm{nat}$C sample per nominal neutron bunch. Symbols show the results of the standard approach for different thresholds on the energy deposited in the C$_{6}$D$_{6}$. The lines represent the results of the Pulse Height Weighting Technique with 200~keV threshold. The values are 68.03$\pm$0.66 and 68.74$\pm$0.44 for the Bicron and FZK detectors, respectively.}
\end{figure}

The second method relies on the use of the Pulse Height Weighting Technique \cite{phwt}. It consists in weighting each count by a suitable factor, determined as a function of the energy deposited in the detectors, to make the efficiency for detecting a capture event independent on the $\gamma$-ray cascade path. As a consequence, capture events can be reliably simulated regardless of the accuracy of the generated cascade or, equivalently, of the $\gamma$-ray spectrum. This is also valid for the background, which mostly originates from capture of scattered neutrons around the sample. For this reason, once the PHWT is applied, the simulated background can reliably be subtracted from the data even for a low threshold on the energy deposited, in this case 200 keV. As a further validation of the technique, it was found that simulations closely reproduce the measured background for time-of-flights below 1.3~ms (corresponding to reconstructed neutron energies above 100~eV), where the contribution from ($n,p$) reactions is negligible. On the other hand, at larger time-of-flight, events from the $^{12}$C($n,p$)$^{12}$B reaction dominate, being up to a factor of six above the background \cite{petarNIM}. Finally, it should be considered that the weighting technique modifies the efficiency of the setup to $^{12}$B-e$^{-}$. The new value, determined from simulations, was used in the analysis.

Figure \ref{fig:two} shows a fit of the measured time-distribution of signals in one of the detectors after background subtraction, when the PHWT is applied, compared with the original efficiency-corrected data for an amplitude threshold of 2~MeV. The results of a pure exponential fit, with 20.2~ms half-life are also shown in the figure for both cases. The reduced chi-square of the fit is in all cases below 1.5. The agreement between the fit and the experimental data, in the whole time range covering four half-lives, provides confidence on the negligible level of the residual background. Similar results are obtained for the other detector.
A more complete view of the results is shown in Fig.~\ref{fig:three}. The ordinate represents the number of ($n,p$) reactions per nominal n\_TOF neutron bunch, as reconstructed from the fit of the time distribution. The symbols show the results for the two detectors as a function of the threshold on the amplitude distribution. Below 1.5~MeV, the presence of a non-negligible residual background leads to an overestimation of the number of ($n,p$) reactions. Above this value, the results are stable against further change in the threshold, all the way up to 3.5~MeV. Since the threshold affects both efficiency corrections and background rejection independently, the stability of the results indicates that both effects are correctly accounted for in the analysis. The straight lines in the figure indicate the results obtained with the Pulse Height Weighting Technique, for a threshold of 200~keV. The agreement between the two different techniques provides a high level of confidence on the validity of the results. From the comparison, an uncertainty of 6\% can be inferred on the reconstructed number of $^{12}$B isotopes produced in this measurement in each neutron bunch.

An additional background contribution is related to elastically scattered neutrons inducing the ($n,p$) reaction in the C$_{6}$D$_{6}$ scintillator itself, as well as in other C-containing material inside the experimental area. Simulations indicate that thanks to the backward position of the detectors such a contribution is less than a percent. It has been subtracted from the data, with a conservative 2\% uncertainty assigned to it. Finally, the present data include a contribution from neutron-induced reactions on $^{13}$C (whose natural abundance is 1.1\%), producing $^{12}$B and $^{13}$B (the latter having similary decay properties of the former). The cross sections of the ($n,p$), ($n,d$) and ($n,np$) reactions on $^{13}$C are highly uncertain, so that no attempt has been made to subtract their contribution from the present data. A realistic 3\% uncertainty has been assigned to the present result, to account for this contribution. The number of produced $^{12}$B isotopes per n\_TOF neutron bunch is 68.5$\pm$0.4(stat)$\pm$4.8(syst).

The activation result reported here represents an integral measurement, with the cross section averaged over the neutron energy spectrum of n\_TOF. As such, it does not allow one to discriminate between different model predictions of the cross section as a function of the neutron energy. Nevertheless, this result can serve as an important constraint for the energy-dependent cross sections obtained via model calculations or evaluations, provided that these are folded with the n\_TOF spectrum, and that other experimental effects are taken into account. The number of $^{12}$B isotopes produced per neutron bunch can be written as:
\begin{equation}
N_{^{12}\mathrm{B}} =  \int_{13.6\:\mathrm{MeV}}^{10\:\mathrm{GeV}} \! \frac{1-e^{-n \sigma_{T}(E)}}{\sigma_{T}(E)} \eta(E)\phi(E)\sigma(E) \, dE
\label{eq:one}
\end{equation}
where $\it{n}$ is the number of atoms/barn of $^{12}$C in the sample, $\phi$ is the neutron flux per unit energy and per bunch, $\sigma$ and $\sigma_{T}$ are the ($n,p$) and total cross section, respectively \cite{supplmat}. Considering that the sample is relatively thick, a correction for multiple scattering, indicated by $\eta$ in the equation, has to be considered. This has been determined from simulations, and can be as high as 14\%. The product of the flux, self-shielding factor and multiple-scattering corrections can be determined as a function of the neutron energy from the simulations. In the equation, it can be replaced by a unique function $\it{w}$(E), that has been fitted with a 5$^{th}$-order polynomial: $\log_{10}[w(E)/w_0]=\sum_{m=0}^5 a_m[\log_{10}(E/E_0)]^m$, where $E_0~=~1$~MeV and $w_0=1$ MeV$^{-1}$mb$^{-1}$, with: $a_0=10.2$, $a_1=-27.5$, $a_2=26.3$, $a_3=-12.3$, $a_4=2.73$ and $a_5=-0.232$.

\begin{figure}[b!]
\includegraphics[angle=0,width=1.\linewidth,keepaspectratio]{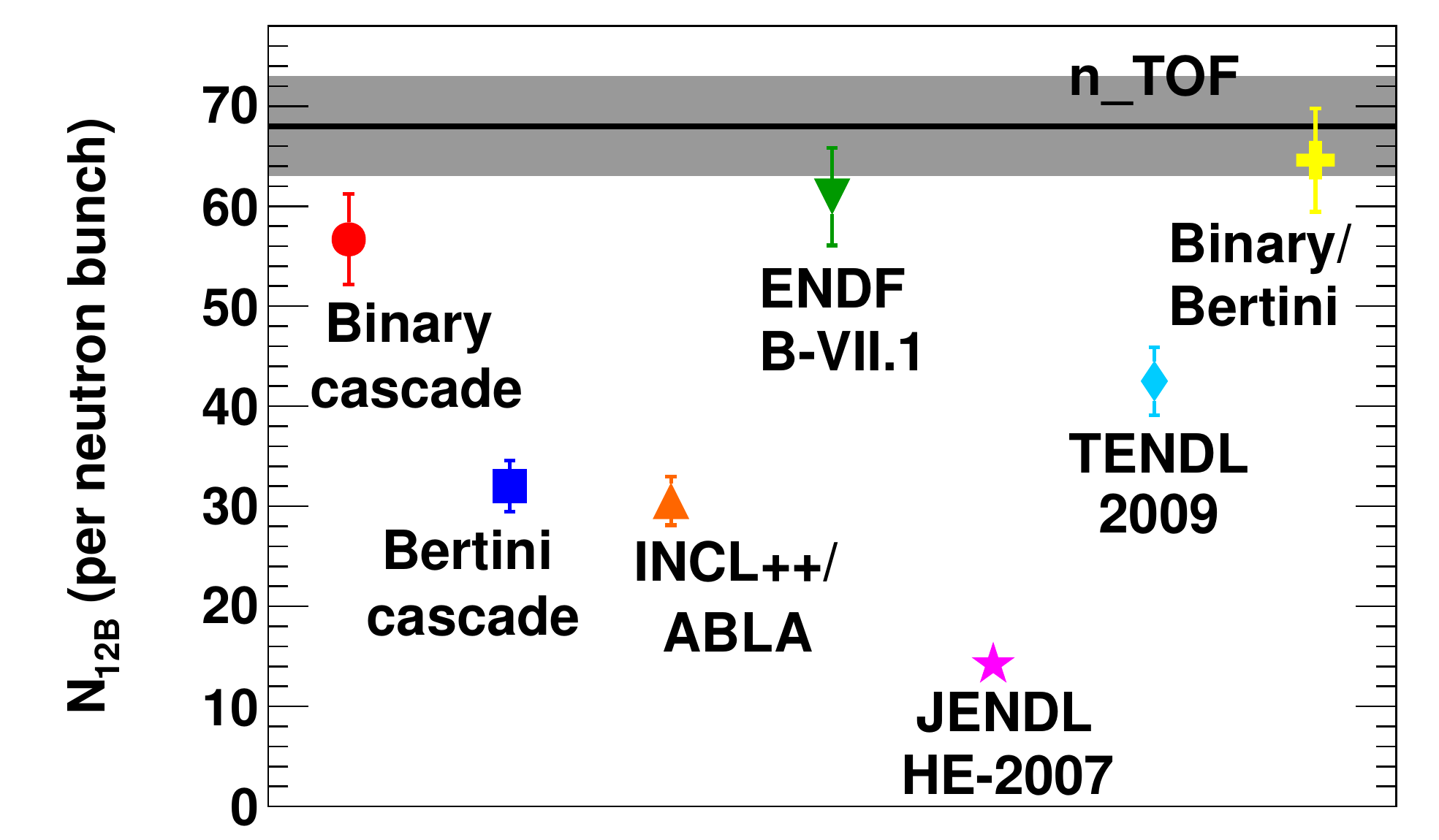}
\caption{\label{fig:four} (Color online) Number of produced $^{12}$B nuclei measured at n\_TOF compared with the one calculated with Eq. (1) for various models and evaluations.}
\end{figure}

The number of $^{12}$B nuclei calculated from Eq. (1) for various model predictions and evaluations is shown in Fig.~\ref{fig:four}. The associated uncertainties are essentially related to the corrections for self-shielding and multiple scattering (5\%) and to the n\_TOF neutron flux (6\%). 

The comparison indicates that JENDL evaluation heavily underestimates the cross section, by almost a factor of five. On the other hand, ENDF/B-VII evaluation is compatible with the present result. However, the presence of a discontinuity in the energy-dependence of the cross section at 20 MeV, evident in Fig.~\ref{fig:one}, indicates that a revision of this library is also in order.  Among models used in GEANT4, a relatively good agreement is observed for the Binary cascade model, while the Bertini cascade and INCL++/ABLA code are off by more than a factor of two. The best result, within 5\% of the experimental value, is obtained by combining the Binary and the Bertini cascade models, with the former used up to 30 MeV and the latter above this energy.

While Eq. (1) should be used for an accurate comparison of the cross section, a less rigorous, but simpler and more general approach can be followed by considering that the product of self-shielding and multiple-scattering corrections is constant within a few percent, and that the n\_TOF neutron flux above 10 MeV is roughly inversely proportional to the neutron energy. In this case, Eq. \ref{eq:one} reduces to: $N_{^{12}\mathrm{B}} \approx c \int \! (\sigma(E)/E)dE$. Comparing the results of this expression with the exact integration of Eq. \ref{eq:one} allows one to determine the proportionality factor \textit{c}. This is found to be constant, within $\pm$5\%, for all employed model, and can therefore be used to extract the experimental value of $\int \! (\sigma(E)/E)dE$=36$\pm$5 mb. This quantity is independent of the specific experimental conditions, in particular sample-related effects and neutron flux, and can therefore be conveniently used for preliminary comparison with model predictions.

In conclusion, we have reported the first measurement of the integral cross section of the $^{12}$C($n,p$)$^{12}$B reaction, performed at n\_TOF from the reaction threshold up to several GeV. The $\beta$-decay of $^{12}$B is detected within the same neutron bunch of the producing reaction. The results indicate that current evaluations are mostly inadequate. In particular, the presence of large discrepancies in the energy-dependence of the cross section calls for further theroretical and experimental efforts to study this reaction. In this respect, the present results may constitute a benchmark for checking the validity of calculations of this reaction cross section, or to tune them for a higher accuracy. The $\it{in-beam}$ activation technique here described can be used for measurements of other reactions of similar features, i.e. leading to the production of a $\beta^{-}$-emitter of millisecond half life, starting from the practically unknown reactions on $^{13}$C.

\begin{acknowledgments}
The research leading to these results has received funding from the European Atomic Energy Community’s (Euratom) Seventh Framework Programme FP7/2007-2011 under the Project CHANDA (GA n. 605203), and by the Croatian Science Foundation under the project 1680. GEANT4 simulations have been run at the Laboratory for Advanced Computing, Faculty of Science, University of Zagreb.
\end{acknowledgments}

\bibliography{C12np}% Produces the bibliography via BibTeX.

\end{document}